\pdfoutput=1
\RequirePackage{ifpdf}
\ifpdf 
\documentclass[pdftex]{sigma}
\else
\documentclass{sigma}
\fi

\usepackage{tabmac}

\newcommand{\cercle}[1]{\ensuremath{\setlength{\unitlength}{1ex}\begin{picture}(2.8,2.8)\put(1.4,0.7){\circle{2.8}\makebox(-5.6,0){#1}}\end{picture}}}
\newcommand{\tcercle}[1]{\ensuremath{\setlength{\unitlength}{1ex}\begin{picture}(2.8,2.8)\put(1.4,1.4){\circle{2.8}\makebox(-5.6,0){#1}}\end{picture}}}

\numberwithin{equation}{section}
\numberwithin{proposition}{section}
\numberwithin{example}{section}
\numberwithin{conjecture}{section}
\numberwithin{definition}{section}
{ \theoremstyle{definition}
\newtheorem{arule}{Rule}[section]}

\DeclareMathOperator{\inv}{inv}

\DeclareMathOperator{\rred}{red}

\begin{document}

\def\cb#1{{\color{blue}#1}} \def\cg#1{{\color{gray}#1}} \def\co#1{{\color{red}#1}} \def\coo#1{{\color{orange}#1}}

\newcommand{\arXivNumber}{1408.2807}

\allowdisplaybreaks

\renewcommand{\PaperNumber}{021}

\FirstPageHeading

\ShortArticleName{Schur Superpolynomials: Combinatorial Def\/inition and Pieri Rule}

\ArticleName{Schur Superpolynomials: Combinatorial Def\/inition\\
and Pieri Rule}

\Author{Olivier BLONDEAU-FOURNIER and Pierre MATHIEU}
\AuthorNameForHeading{O.~Blondeau-Fournier and P.~Mathieu}
\Address{D\'epartement de physique, de g\'enie physique et d'optique, Universit\'e Laval,\\ Qu\'ebec, Canada, G1V 0A6}

\Email{\href{mailto:olivier.b-fournier.1@ulaval.ca}{olivier.b-fournier.1@ulaval.ca},
\href{mailto:pmathieu@phy.ulaval.ca}{pmathieu@phy.ulaval.ca}}

\ArticleDates{Received August 28, 2014, in f\/inal form February 25, 2015; Published online March 11, 2015}

\Abstract{Schur superpolynomials have been introduced recently as limiting cases of the Macdonald superpolynomials.
It turns out that there are two natural super-extensions of the Schur polynomials: in the limit $q=t=0$ and
$q=t\rightarrow\infty$, corresponding respectively to the Schur superpolynomials and their dual.
However, a~direct def\/inition is missing.
Here, we present a~conjectural combinatorial def\/inition for both of them, each being formulated in terms of a~distinct
extension of semi-standard tableaux.
These two formulations are linked by another conjectural result, the Pieri rule for the Schur superpolynomials.
Indeed, and this is an interesting novelty of the super case, the successive insertions of rows governed by this Pieri
rule do not generate the tableaux underlying the Schur superpolynomials combinatorial construction, but rather those
pertaining to their dual versions.
As an aside, we present various extensions of the Schur bilinear identity.}


\Classification{05E05}

\section{Introduction}

\subsection{Schur polynomials}

The simplest def\/inition of the ubiquitous Schur polynomials is combinatorial~\cite[I.3]{Mac} (and also~\cite[Chapter~4]{Sag}).
For a~partition~$\lambda$, the Schur polynomial $s_\lambda(x)$, where~$x$ stands for the set $(x_1,\ldots, x_N)$, is
the sum over monomials weighted by the content of semi-standard tableaux~$T$ with shape~$\lambda$ (whose set is denoted
${\mathcal T}(\lambda)$):
\begin{gather}
s_\lambda(x)=\sum\limits_{T\in{\mathcal T}(\lambda)}x^T,
\label{sT}
\end{gather}
with $x^T\equiv x^\alpha=x_1^{\alpha_1}x_2^{\alpha_2}\cdots x_N^{\alpha_N}$, where~$\alpha$ is a~composition of weight
$|\lambda|$ (the sum of all the parts of~$\lambda$) and represents the content of~$T$, namely~$T$ contains $\alpha_i$ copies of~$i$.
For instance, in three variables,
\begin{gather*}
s_{(2,1)}(x)=x_1^2x_2+x_1x_2^2+x_1^2x_3+x_1x_3^2+x_1x_2x_3+x_1x_2x_3,
\end{gather*}
where each term is in correspondence with a~semi-standard tableau:
\[
\tableau[scY]{1&1\\2},\qquad\tableau[scY]{1&2\\2},\qquad\tableau[scY]{1&1\\3},\qquad\tableau[scY]{1&3\\3},\qquad\tableau[scY]{1&2\\3},
\qquad\tableau[scY]{1&3\\2}.
\]
The number of semi-standard tableaux of shape~$\lambda$ and content~$\mu$ (where~$\mu$ is also a~partition) is the
Kostka number $K_{\lambda\mu}$.
Relation~\eqref{sT} can thus be rewritten
\begin{gather*}
s_\lambda(x)={\sum\limits_{\alpha}}K_{\lambda\mu} x^\alpha=\sum\limits_{\mu\leq\lambda}K_{\lambda\mu} m_\mu(x),
\end{gather*}
where $\mu=\alpha^+$, the partition obtained by ordering the parts of the composition~$\alpha$ in weakly decreasing order.
In the last equality, the symmetric character of the Schur polynomial is used to reexpress them in terms of
the monomial symmetric polynomials $m_\mu(x)$,
\begin{gather}
\label{mono}
m_\mu(x)=x_1^{\mu_1}\cdots x_N^{\mu_N}+\text{distinct permutations},
\end{gather}
and the order in the summation is the dominance order: $\lambda\geq \mu$ if and only if $|\lambda|=|\mu|$ and
$\sum\limits_{i=1}^k(\lambda_i-\mu_i)\geq 0$ for all~$k$.

A closely related approach for constructing the Schur polynomials is based on the Pieri formula.
The latter refers to the decomposition in Schur polynomials of the product of two Schur polynomials, one of which being
indexed by a~single-row partition, say $s_{(k)}$.
Schur polynomials can thus be constructed from the identity by multiplying successive $s_{(k)}$'s.
Recall that $s_{(k)}=h_k$, where $h_k$ stands for the completely homogeneous symmetric polynomials:
\begin{gather*}
h_k(x)=\sum\limits_{1\leq i_1 \leq \cdots \leq i_k\leq N} x_{i_1}\cdots x_{i_k},
\end{gather*}
to which is associated the multiplicative basis $h_\mu=h_{\mu_1}h_{\mu_2}\cdots h_{\mu_N}$ (with $h_0=1$).
The link between the Pieri rule and the above def\/inition of the Schur polynomials is
\begin{gather*}
s_{(\mu_1)}s_{(\mu_2)}\cdots s_{(\mu_N)}=h_{\mu_1}h_{\mu_2}\cdots h_{\mu_N}=h_\mu(x)=\sum\limits_{\lambda\geq\mu}K_{\lambda\mu} s_\lambda(x).
\end{gather*}
In other words, by f\/illing the rows $\mu_i$ with number~$i$ and applying the Pieri rule from left to right to evaluate
the multiple product of the $s_{(\mu_i)}$, amounts to construct semi-standard tableaux.
Then, isolating the term $s_\lambda$ from the complete product is equivalent to enumerate the semi-standard tableaux of
shape~$\lambda$ and content~$\mu$.

The objective of this work is to present conjectured combinatorial def\/initions of the Schur superpolynomials and their
dual.
In addition, we present the conjectural version of the Pieri rule for Schur superpolynomials.
{By duality}, the resulting tableaux are not those appearing in the combinatorial description of the Schur
superpolynomials but rather {those} of their dual version.

\subsection{Schur superpolynomials}

Superpolynomials refer to polynomials in commuting and anticommuting variables (respectively denoted~$x_i$ and
$\theta_i$) and their symmetric form entails invariance under the simultaneous permutation of the two types of variables.
The classical symmetric functions are generalized in~\cite{DLMclass}.
Super-analogues of the Jack and Macdonald symmetric polynomials appear as eigenfunctions of the Hamiltonian of the
supersymmetric extension of the Calogero--Sutherland (cf.~\cite{DLM1,DLM3}) and the Ruijsenaars--Schneider
(cf.~\cite{BDM}) models respectively.

Denote the Macdonald superpolynomials by $P_\Lambda{=} P_\Lambda(x,\theta;q,t)$, where~$\Lambda$ is a~superpartition
(cf.~Section~\ref{Spart}), $(x,\theta)$ denotes the $2N$ variables ${(x_1,\ldots, x_N, \theta_1,\ldots,\theta_N)}$ and
$q$, $t$ are {two arbitrary} parameters.
$P_\Lambda$ is def\/ined from~\cite{BDLM1}:
\begin{gather*}
1)
\
P_\Lambda=m_\Lambda+\text{lower terms},
\\
2)
\
\langle P_\Lambda, P_\Omega \rangle_{q,t} \propto \delta_{\Lambda \Omega},
\end{gather*}
where lower terms are w.r.t.~the dominance order between superpartitions (cf.~Def\/inition~\ref{dom}), $m_\Omega$ {stands for
the super-monomial} (cf.~Def\/inition~\ref{SM}) and the scalar product is def\/ined in terms of the power-sum basis $p_\Lambda$
(cf.~Def\/inition~\ref{MulSB}, equation~\eqref{Spower}) with
\begin{gather}
\label{psqt}
\langle p_\Lambda, p_\Omega \rangle_{q,t} {:=} \delta_{\Lambda \Omega}(-1)^{\binom{m}{2}} z_\Lambda(q,t),
\qquad
z_\Lambda(q,t) {=} z_{\Lambda^s} q^{|\Lambda^a|}\prod\limits_i \frac{1-q^{\Lambda^s_i}}{1-t^{\Lambda^s_i}},
\end{gather}
where we used the representation $\Lambda=(\Lambda^a;\Lambda^s)$ with~$m$ being the number of parts of $\Lambda^a$, and
$z_\lambda=\prod\limits_i i^{n_i(\lambda)}n_i(\lambda)!$ where $n_i(\lambda)$ is the number of parts equal to~$i$
in~$\lambda$.
Two special {limits} of the~$P_\Lambda$ are of particular interest, namely
\begin{gather}
\label{2ss}
s_\Lambda(x,\theta):=P_\Lambda(x,\theta;0,0)
\qquad
\text{and}
\qquad
\bar s_\Lambda(x,\theta):=P_\Lambda(x,\theta;\infty,\infty).
\end{gather}
That $s_\Lambda$ and $\bar s_\Lambda$ are dif\/ferent objects follows from the fact that
\begin{gather*}
P_\Lambda(x,\theta;q,t)\ne P_\Lambda\big(x,\theta;q^{-1},t^{-1}\big)
\qquad
\text{since}
\qquad
z_\Lambda\big(q^{-1},t^{-1}\big) \neq f(q,t)^{|\Lambda|}z_\Lambda(q,t)
\end{gather*}
for some {monomial} $f(q,t)$.
We refer to the functions $s_\Lambda$ and $\bar s_\Lambda$ as the Schur superpolynomials and the dual Schur
superpolynomials respectively.

A closer look at~\eqref{2ss} shows that these are not {at once} sound def\/initions since in both limits, the scalar
product~\eqref{psqt} is ill-def\/ined.
However, a~deeper investigation relying on~\cite{FL,Ion2} reveals that both limiting forms of the Macdonald
superpolynomials turn out to be well behaved and this results in the expression of $s_\Lambda$ and $\bar s_\Lambda$ in
terms of key polynomials (see~\cite[Appendix~A]{BDLM2}).

Here we present an alternative and direct~-- albeit conjectural~-- combinatorial def\/inition for both $s_\Lambda$ and $\bar s_\Lambda$.
It relies on the observation that the elements of the transition matrices between the $s_\Lambda$, $\bar s_\Lambda$
and the super-monomial basis are all non-negative integers~\cite[Conjecture~6]{BDLM1}. In other words, we have that
\begin{gather*}
s_\Lambda=\sum\limits_{\Lambda \geq \Omega} \bar{K}_{\Lambda \Omega} m_\Omega,
\qquad
\bar{s}_\Lambda=\sum\limits_{\Lambda \geq \Omega} {K}_{\Lambda \Omega} m_\Omega
\end{gather*}
with $\bar{K}_{\Lambda \Omega}$, ${K}_{\Lambda \Omega} \in \mathbb{N}$.\footnote{The somewhat
unnatural position of the bar over ${K}_{\Lambda \Omega}$ with respect to that over $s_\Lambda$ follows the notation of~\cite{BDLM1}.
The functions $s_\Lambda$ are used to def\/ine a~novel positivity conjecture~\cite{Hai} in superspace (see~\cite[Conjecture~9]{BDLM1}).
If we denote by $J_\Lambda$ the integral version of the Macdonald superpolynomials ($J_\Lambda=c_\Lambda P_\Lambda$ with
$c_\Lambda\in \mathbb{Z}(q,t)$) and $S_\Lambda(t)=\varphi(s_\Lambda)$, where $\varphi(\cdot)$ is the homomorphism that
send $p_r \rightarrow (1-t^r)p_r$, we obtain that
\begin{gather*}
J_\Lambda=\sum\limits_\Omega K_{\Omega \Lambda}(q,t) S_\Omega(t),
\qquad
K_{\Omega \Lambda}(q,t)\in \mathbb{N}(q,t).
\end{gather*}
Setting $q=0$, $t=1$ in $K_{\Omega \Lambda}(q,t)$ gives the number $K_{\Omega \Lambda}$ that appears in the super-monomial
expansion of $\bar{s}_\Omega$ (not~$s_\Omega$).
Note that there is no relation between the numbers $\bar{K}_{\Lambda \Omega}$ and ${K}_{\Lambda \Omega}$ except that
they both reduce to ordinary Kostka numbers for $m=0$.}
The integral positivity of these coef\/f\/icients hints for an
underlying combinatorial description {for both $s_\Lambda$ and $\bar s_\Lambda$.
The presentation of such a~combinatorial description is the main result of this work.}   The expansion coef\/f\/icients
$\bar{K}_{\Lambda \Omega}$ and ${K}_{\Lambda \Omega}$ are obtained by enumerating appropriate generalizations of
semi-standard tableaux.
Despite the fact that there is no known representation theory underlying these new
generalizations of the Schur functions, we expect that the present tableau construction will pave the way for the
elaboration of a~super-version of the Robinson--Schensted--Knuth correspondence.

The article is organized as follows.
We f\/irst introduce the notion of super semi-standard tableaux in Section~\ref{SST}.
Those tableaux of shape~$\Lambda$ are the building blocks for def\/ining the \emph{combinatorial} Schur superpolynomial
$s^c_\Lambda$, presented in Section~\ref{sSchur}.
The construction readily implies the symmetric nature of $s^c_\Lambda$ and
its triangular decomposition in the monomial {basis}.
The Pieri rule for the $s_\Lambda$ is given in Section~\ref{Pieri}.
In Section~\ref{dual} we introduce the dual super semi-standard tableaux whose enumeration describes (conjecturally) the dual
version~$\bar s_\Lambda$.
The link with Pieri tableaux that arise in the multiple application of the Pieri rule for Schur superpolynomials is
spelled out in Section~\ref{Section6}.
Two appendices complete this article.
The f\/irst is a~review of the necessary tools concerning superpartitions and the classical bases in superspace.
In Appendix~\ref{Bil}, we present a~collection of generalized bilinear identities for Schur superpolynomials.

\section{Super semi-standard tableaux} \label{SST}

In this section, we generalize semi-standard tableaux to diagrams associated with superpartitions.
By a~super tableau, we refer to the f\/illing of all the boxes and the circles of the diagram of a~superpartition with
numbers from a~given set.
A~super semi-standard tableau $T^\circ$ of shape~$\Lambda$ of degree $(n|m)$
(cf.~{Section~\ref{Spart}}), is a~f\/illing of each of the~$n$ boxes and~$m$ circles in the diagram of~$\Lambda$ with
{numbers} from the set $I=\{1, 2, \ldots, N\}$, for $N\geq n$, {and} subject to the following rules.
(The $\circ$ upper-script in $T^\circ$ reminds that these are tableaux containing circles.)

\begin{arule}[numbers in circles]\label{numcirc}
The numbers in the~$m$ circles of $T^\circ$ are~$m$ distinct numbers from the set~$I$.
\end{arule}

The circled numbers are called fermionic numbers and they form the ordered set $I_m=(i_1, \ldots, i_m)$ where the circle
content is read from top to bottom.
Numbers in the complementary set $I_m^c:=I\setminus I_m$ are called bosonic numbers.

\begin{arule}[ordering in the set~$I$]\label{order}
The numbers in the set $I_m$ are considered to be the~$m$ largest numbers of the set~$I$ and are ordered as
$i_1>i_2>\cdots >i_m$.
In addition $i_m>j$ $\forall\, j\in I_m^c$.
In $I_m^c$, the ordering is the natural one.
\end{arule}

For example, for the following circle f\/illing:
\begin{equation}
\label{exfill}
\tableau[scY]{&&&&&&\bl\tcercle{3}\\&&&&\bl\tcercle{1}\\&&\\&\bl\tcercle{5}},
\end{equation}
$I_3=(3,1,5)$ and, with $N=14$, the ordering in~$I$ reads: $3>1>5>14>13>\dots>6>4>2$.

\begin{arule}[numbers in boxes of fermionic rows]
The numbers in the boxes of each fermionic row of $T^\circ$ are all identical and f\/ixed to be equal to the number of the
ending circle.
These numbers are said to be frozen and play a~passive role in regard to constraints on the numbers placed in other
boxes (in particular, with respect to the number ordering in columns).
\end{arule}

For the above example, we have
\[
\tableau[scY]{3&3&3&3&3&3&\bl\tcercle{3}\\1&1&1&1&\bl\tcercle{1}\\&&\\5&\bl\tcercle{5}}.
\]

We now introduce additional rules for fermionic numbers in bosonic rows.
We f\/irst need the following def\/inition.

\begin{definition}[fermionic singlets and doublets]
Let $i_k$ and $i_{k+1}$ be two consecutive fermionic numbers, $k\in\{1, \ldots, m-1\}$, lying in distinct bosonic rows.
A~pair $\{i_{k+1},i_k\}$ is said to form a~doublet if $i_k$ lies in a~row lower than the $i_{k+1}$.
Other occurrences of $i_{k+1}$ are said to be singlets.
\end{definition}

 Consider, for example, the following partially f\/illed tableaux (where only fermionic numbers are considered):
\begin{equation}
\label{exfixedpair}
\tableau[scY]{
&&&&2&2\\&&&2&1  \\ 1&1&1 &\bl \tcercle{1} \\ & \\ \bl\tcercle{2}
}, \qquad
\tableau[scY]{
&&&&&\\&&2&2&1  \\ 1&1&1 &\bl \tcercle{1} \\ &2 \\ \bl\tcercle{2}
}.
\end{equation}
For the tableau on the left, the number $2$ in box $(1,5)$ belongs to a~doublet since it can be coupled with the~1 in
position $(2,5)$; the remaining two $2$ are singlets.
(The~2 in the doublet could as well be taken to be the one in position $(1,6)$.)
Note that the three~1 in the third row cannot be parts of doublets being in a~fermionic row.
In the second tableau, the three $2$ are singlets.
As we just remarked, the characterization of a~doublet is not unique.
However, their number is well def\/ined and this implies that the number of singlets is also well def\/ined.
It is precisely the number of singlets that is relevant here.

\begin{arule}[fermionic numbers in bosonic rows]
Let   $c_{k,0}$ and $c_{k,k-1}$ be respectively the number of bosonic columns at the right of the fermionic column
ending with a~circled $i_k$ and the number of bosonic columns between the two fermionic columns ending with circled
$i_k$ and $i_{k-1}$.
The occurrence of $i_k$ in boxes of bosonic rows must satisfy the two conditions:
\begin{enumerate}\itemsep=0pt
\item[$(a)$] $i_k$ can only appear at the (upper) right of the fermionic column ending with circle $i_k$;

\item[$(b)$] $i_k$ can appear at most $c_{k,0}$ times but at most $c_{k,k-1}$ times as singlet.
\end{enumerate}
\end{arule}

For instance, the tableau at the right in~\eqref{exfixedpair} does not satisfy this rule since there can be at
most two singlets~2 given that $c_{2,1}=2$.
As a~further example, notice that
\[
\tableau[scY]{&&&\\&&&\\&&&3\\&&\bl\tcercle{1}\\&\bl\tcercle{2}\\\bl\tcercle{3}}\qquad\text{and}\qquad
\tableau[scY]{&&&\\&&&3\\&&&2\\&&\bl\tcercle{1}\\&\bl\tcercle{2}\\\bl\tcercle{3}}
\qquad\text{are not allowed but}\qquad
\tableau[scY]{&&&3\\&&&2\\&&&1\\&&\bl\tcercle{1}\\&\bl\tcercle{2}\\\bl\tcercle{3}}\quad\text{is allowed.}
\]
Indeed, since $c_{3,2}=0$, the~3 cannot be a~singlet.
In the second tableau the~3 is part of a~doublet but now $c_{2,1}=0$ requires the~2 to be coupled with~1.

\begin{arule}[semi-standard f\/illing]\label{SSfil}
The bosonic rows are f\/illed with numbers in~$I$ using the ordering def\/ined in Rule~\ref{order} and such that numbers in
rows are weakly increasing and strictly increasing in columns (disregarding the frozen numbers in fermionic rows).
\end{arule}

\begin{definition}
A~tableau $T^\circ$ satisfying Rules~\ref{numcirc}--\ref{SSfil} is called a~super semi-standard tableau.
\end{definition}

Here is an example
\begin{equation}
\label{ex}
\tableau[scY]{
1&1&1&1&1&1&1&1&\bl\tcercle{$1$}\\2&2&2&2&2&2&\bl\tcercle{2}\\6&6&7&8&5\\3&3&3 & \bl\tcercle{3}\\7&7&8\\4&4&\bl\tcercle{4}\\\bl\tcercle{5}
}.
\end{equation}

When fermionic numbers appear in bosonic rows, being larger than the bosonic numbers, they occupy the rightmost
positions (and by consequence, the downmost positions in columns).
Once all fermionic numbers have been inserted, it is convenient to work with a~reduced diagram, or, if f\/illed, a~reduced tableau.

\begin{definition}[reduced tableau]
The reduced tableau of $T^\circ$, denoted $T^\circ_{\rred}$, is the tableau obtained from $T^\circ$ by removing
all circles and all the boxes marked by fermionic numbers and tighten all rows.
\end{definition}

The numbers in boxes of the reduced tableau satisfy the ordinary semi-standard tableau conditions for the numbers in the set~$I_m^c$.
In other words, $T^\circ_{\rred}$ is a~genuine semi-standard tableau.
For the example~\eqref{ex}, $T^\circ_{\rred}$ is simply
\[
\tableau[scY]{6&6&7&8\\7&7&8}.
\]

\section{The combinatorial Schur superpolynomials}\label{sSchur}

In this section, we introduce a~new family of functions, dubbed the combinatorial Schur superpolynomials~$s^c_\Lambda$.
They are def\/ined in terms of the tableaux introduced in the previous section.
That these functions are indeed equal to the Schur superpolynomials def\/ined in~\eqref{2ss} is the main (conjectural)
result of this section.

Let~$\Lambda$ be of degree $(n|m)$.
We def\/ined a~monomial $\zeta^{T^\circ}$ in the variables $(x,\theta)$ associated to the tableau $T^\circ$ by introducing
a~factor $x_i$ for each number~$i$ appearing in a~box and a~$\theta_j$ for a~circled~$j$, the circle content being read
from top to bottom:
\begin{gather*}
\zeta^{T^\circ}:=\theta_{I_m} \prod\limits_{i\in T^*} x_i
\end{gather*}
with $\theta_{I_m}=\theta_ {i_1} \cdots \theta_{i_m}$ and $T^*$ denotes the box content of $T^\circ$.
For instance, the monomial corresponding to the tableau~\eqref{ex} is
\begin{gather*}
\theta_1\theta_2\theta_3\theta_4\theta_5x_1^8x_2^6x_3^3x_4^2x_5x_6^2x_7^3x_8^2.
\end{gather*}

\begin{definition}
The combinatorial Schur superpolynomial $s^c_\Lambda=s^c_\Lambda(x,\theta)$ is given~by
\begin{gather}
\label{SenT}
s^c_\Lambda:=\sum\limits_{{T^\circ} \in{\mathcal T}^\circ(\Lambda)} \zeta^{T^\circ},
\end{gather}
where ${\mathcal T}^\circ(\Lambda)$ denotes the set of super semi-standard tableaux of shape~$\Lambda$.
\end{definition}

The upper script~$c$ refer to the combinatorial def\/inition: $s^c_\Lambda$ might dif\/fer in principle from
$s_\Lambda$ def\/ined in~\eqref{2ss}.

For example, $s^c_{(0;2,1)}$ is obtained by summing the contribution of the tableaux:
\[
\tableau[scY]{2&2 \\3 \\ \bl\tcercle{1} }, \!\quad
\tableau[scY]{2&3 \\3 \\ \bl\tcercle{1} }, \!\quad
\tableau[scY]{1&1 \\3 \\ \bl\tcercle{2} }, \!\quad
\tableau[scY]{1&3 \\3 \\ \bl\tcercle{2} }, \!\quad \dots  , \!\quad
\tableau[scY]{2&1 \\3 \\ \bl\tcercle{1} }, \!\quad
\tableau[scY]{1&2 \\3 \\ \bl\tcercle{2} }, \!\quad \dots  , \!\quad
\tableau[scY]{2&3 \\4 \\ \bl\tcercle{1} }, \!\quad
\tableau[scY]{2&4 \\3 \\ \bl\tcercle{1} }, \!\quad \dots  ,
\]
whose variable transcription reads
\begin{gather*}
\theta_1x_2^2x_3+\theta_1x_2x_3^2+\theta_2 x_1^2x_3+\theta_2x_1x_3^2+\cdots+\theta_1x_1x_2 x_3+\theta_2x_1x_2
x_3+\cdots+2 \theta_1 x_2x_3 x_4+\cdots.
\end{gather*}
In order to show that the superpolynomials $s^c_\Lambda$ can be written in the basis of the monomial superpolynomials,
their symmetric character must f\/irst be established.

\begin{proposition}
\label{sisi}
The superpolynomial $s^c_\Lambda$ is symmetric.
\end{proposition}

\begin{proof}
It suf\/f\/ices to show that $s^c_\Lambda$ is invariant under elementary transpositions, {e.g.:}
\begin{gather*}
({i,i+1}) s^c_\Lambda=s^c_\Lambda,
\end{gather*}
where $(i,i+1)$ is the permutation that exchanges simultaneously $x_{i} \leftrightarrow x_{i+1}$ and $\theta_{i}\leftrightarrow \theta_{i+1}$.
This action is lifted to tableaux as follows.
Consider the involution: $T^\circ\rightarrow \tilde T^\circ$ such that the number of boxes marked with~$i$'s and the
number of boxes marked with $(i+1)$'s are exchanged from $T^\circ$ to $\tilde T^\circ$.
Two cases need to be described.

(1) If~$i$ and $i+1$ are bosonic numbers (i.e.~$\in I_m^c$), the involution only transforms the reduced tableaux, which are semi-standard.
In that case, the involution is taken to be the usual one (see, e.g.,~\cite[Proposition~4.4.2]{Sag}).
The tableau reconstructed from the modif\/ied reduced tableau is manifestly an element of~${\mathcal T}^\circ(\Lambda)$.

(2) If~$i$ and/or $i+1 \in I_m$, the operation simply amounts to interchange all the numbers~$i$ and $i+1$ in $T^\circ$.
In that case, the ordering within the set $I_m$ is modif\/ied and the new f\/illing is automatically an element of
${\mathcal T}^\circ(\Lambda)$.
\end{proof}

To substantiate the conjectural equivalence of $s_\Lambda^c$ and $s_\Lambda$ (cf.~Conjecture~\ref{ssc} below), we
demonstrate the unitriangularity of $s_\Lambda^c$ in its super-monomial expansion.

\begin{proposition}
We have
\begin{gather}
\label{sLaenKbarmOm}
s^c_\Lambda=\sum\limits_{\Omega} \bar{K}^c_{\Lambda \Omega} m_\Omega,
\end{gather}
where $\bar{K}^c_{\Lambda \Omega}$ is the number of super semi-standard tableaux of shape~$\Lambda$ and
content~$\Omega$, with
\begin{gather}
\label{Kbartriang}
\bar{K}^c_{\Lambda\Omega}=
\begin{cases}
0&\text{if}\quad \Omega \not \leq \Lambda,
\\
1&\text{if}\quad \Omega=\Lambda.
\end{cases}
\end{gather}
\end{proposition}

\begin{proof}
Since $s^c_\Lambda$ is symmetric, the decomposition~\eqref{SenT} can be rewritten as an expansion in terms of the
monomials $m_\Omega$ (which form a~basis for symmetric superpolynomials~\cite{DLMclass}).
As a~result, the expression~\eqref{sLaenKbarmOm} follows directly from the def\/inition of $\bar K^c_{\Lambda\Omega}$ as
the number of elements of ${\mathcal T}^\circ(\Lambda)$ having content $\Omega$. What has to be shown then is the
statement~\eqref{Kbartriang}, i.e., that the expansion is unitriangular.

At f\/irst, observe that one can focus on tableaux for which $I_m={\{1, \ldots, m\}}$ and with box content
$(1^{\Omega_1},2^{\Omega_2}, \ldots, N^{\Omega_N})$ in order to identify the multiplicity of the monomial $m_\Omega$ since:
\begin{gather*}
m_\Omega=\theta_1 \cdots \theta_m x_1^{\Omega_1} \cdots x_N^{\Omega_N}+\text{distinct permutations}.
\end{gather*}

Now, consider all the dif\/ferent monomials $m_\Omega$ appearing in the expansion of $s^c_\Lambda$,
cf.~equation~\eqref{sLaenKbarmOm}.
First, consider those f\/illings of~$\Lambda$ such that no fermionic numbers appear in bosonic rows.
In this case, the content $\Omega=(\Omega^a;\Omega^s)$ is necessarily such that $\Omega^a=\Lambda^a$.
We are left with the f\/illing of reduced diagrams of shape $\Lambda^s$ and content $\Omega^s$.
In other words, we have
\begin{gather*}
\bar K^c_{(\Lambda^a;\Lambda^s) (\Lambda^a;\Omega^s)}=K_{\Lambda^s,\Omega^s},
\end{gather*}
where $K_{\Lambda^s,\Omega^s}$, being indexed by two ordinary partitions, refers the usual Kostka numbers.
The unitriangularity of the $K_{\Lambda^s,\Omega^s}$'s proves the triangularity of $\bar{K}^c_{\Lambda\Omega}$ for the
special case where \mbox{$\Omega^a=\Lambda^a$} and it implies that $\bar K^c_{\Lambda \Lambda}=1$.

Next, consider the case where $\Omega^a\ne \Lambda^a$.
By construction, we have $\Omega^a>\Lambda^a$ (with respect to the dominance ordering for partitions but relaxing the
constraint $|\Omega^a|=|\Lambda^a|$).
Indeed, the numbers~$k$ within a~row, say~$i$, ending with the circle~$k$ are frozen and some extra~$k$ may be inserted
in the upper-right part of the tableau; therefore $\Omega^a_i\geq \Lambda^a_i$.
Suppose that one fermionic number~$k$ is introduced in row~$i$.
Focussing on the tails of the two rows concerned here, we have:
\[
\underset{\substack{\text{part of $\Lambda$}\\ \text{with  a marked circle}}}{\tableau[scY]{\bl{\!\!\!\cdots}&&&&&&\\ \bl{\!\!\!\cdots}&\bl\tcercle{$k$}}}  \quad \rightarrow \quad \underset{\text{part of content $\Omega$}}{\tableau[scY]{\bl{\!\!\!\cdots}&&&&&&k\\\bl{\!\!\!\cdots}& \bl\tcercle{$k$} } }\quad \rightarrow \quad
\underset{\text{part of shape $\Omega$}}{\tableau[scY]{\bl{\!\!\!\cdots}&&&&&\\\bl{\!\!\!\cdots}&k& \bl\tcercle{$k$} } }\,.
\]
Clearly, the relation between~$\Lambda$ and~$\Omega$ is as follows: the diagram of~$\Omega$ is obtained from that
of~$\Lambda$ by moving a~box downward from a~bosonic to a~fermionic row.
This operation satisf\/ies $\Lambda^* \geq {\Omega}^*$ and $\Lambda^\circledast \geq {\Omega}^\circledast$ so that
$\Lambda \geq \Omega$ (cf.~Def\/inition~\ref{dom}).
Indeed, it suf\/f\/ices to compare the superpartitions composed of the two concerned rows: $(b;a)\in \Lambda$ and $
(b+1;a-1) \in \Omega$, with $a>b$, and testing successively the truncated version of $\Lambda^* \geq {\Omega}^*$ and
$\Lambda^\circledast \geq {\Omega}^\circledast$:
\begin{gather*}
(a,b)\geq (a-1,b+1)
\quad
\text{and}
\quad
(a,b+1)\geq (a-1,b+2)
\quad
\implies
\quad
(b;a)\geq(b+1;a-1).
\end{gather*}
Now consider inserting several~$k$ in the same row (here~3)
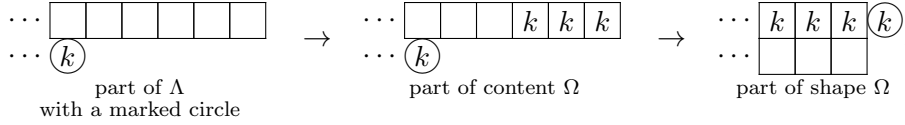
\[
\underset{\substack{\text{part of~$\Lambda$}
\\
\text{with a~marked circle}}}{\tableau[scY]{\bl{\!\!\!\cdots}&&&&&&\\ \bl{\!\!\!\cdots}&\bl\tcercle{$k$}}}
\quad
\rightarrow
\quad
\underset{\text{part of content~$\Omega$}}{\tableau[scY]{\bl{\!\!\!\cdots}&&&&k&k&k\\\bl{\!\!\!\cdots}& \bl\tcercle{$k$}}}
\quad
\rightarrow
\quad
\underset{\text{part of shape~$\Omega$}}{\tableau[scY]{\bl{\!\!\!\cdots}&k&k&k& \bl\tcercle{$k$}\\\bl{\!\!\!\cdots}&&&}}.
\]
In the f\/inal step, at the level of the shapes of the corresponding diagrams, a~box has moved downward from a~bosonic to
a~fermionic row and a~circle has moved up.
A~computation similar to the above one consists in comparing the two-row superpartitions $(b;a)$ and $(a-1;b+1)$ for $a>b$; we see that
\begin{gather*}
(a,b)\geq (a-1,b+1)
\quad
\text{and}
\quad
(a,b+1)\geq (a,b+1)
\quad
\implies
\quad
(b;a)\geq(a-1;b+1)
\end{gather*}
from which it follows that $\Lambda^* \geq \Omega^*$ and $\Lambda^\circledast \geq \Omega^\circledast$.
Since these two processes can be done iteratively, the action of f\/illing several boxes of bosonic rows with fermionic
numbers {always produce terms of lower degree with respect to} the dominance ordering.
\end{proof}

We now turn to some examples illustrating the construction of the combinatorial object $s_\Lambda^c$.
For all the examples, we identify the leading term of a~super-monomial and consider thus only those tableaux $T^\circ_m$
for which all circles are f\/illed with the numbers $(1,\ldots,m)$, from top to bottom.

\begin{example}
The case of a~purely fermionic superpartition.
Here there is a~single contributing monomial:
\begin{gather*}
s^c_{(\Lambda^a; )}=m_{(\Lambda^a; )}.
\end{gather*}
Indeed, there are no bosonic rows so that there is only one contributing tableau $T^\circ_m$ where all the boxes are
frozen, all marked~$i$ in row~$i$.
\end{example}

\begin{example}
The case of a~superpartition where
 $(\Lambda^a,\Lambda^s)$ is a~partition (i.e., $\Lambda_m\geq \Lambda_{m+1}$).
Then we have
\begin{gather*}
s^c_\Lambda=\sum\limits_{\Omega^s \leq \Lambda^s} K_{\Lambda^s,\Omega^s} m_{(\Lambda^a;\Omega^s)}.
\end{gather*}
Here the bosonic rows of $\Lambda^s$ all lie below the fermionic ones.
Since the fermionic number~$k$ cannot appear in rows below the one ending with circled~$k$, there cannot be any
occurrence of the fermionic numbers in the bosonic rows.
These bosonic boxes are then f\/illed with numbers in the set $\{m+1,\dots,N\}$, generating usual semi-standard tableaux
enumerated by the usual Kostka coef\/f\/icients.
\end{example}

\begin{example}
A~simple case: $\Lambda=(1;4)$.
With $I_1=\{1\}$, the f\/illed tableaux are (note that here $1>4>3>2$):
\begin{equation}\label{tabex3}
\begin{split}
&
 {\tableau[scY]{2&2&2&2\\{1}&\bl\tcercle{1}},
\qquad
\tableau[scY]{2&2&2&3\\1&\bl\tcercle{1}},
\qquad
\tableau[scY]{2&2&3&3\\1&\bl\tcercle{1}},
\qquad
\tableau[scY]{2&2&3&4\\1&\bl\tcercle{1}},
\qquad
\tableau[scY]{2&3&4&5\\1&\bl\tcercle{1}}},
\\
& { \tableau[scY]{2&2&2&1\\1&\bl\tcercle{1}},
\qquad
\tableau[scY]{2&2&3&1\\   1&\bl\tcercle{1}},
\qquad
\tableau[scY]{2&3&4&1\\   1&\bl\tcercle{1}},
\qquad
\tableau[scY]{2&2&1&1\\   1&\bl\tcercle{1}},
\qquad
\tableau[scY]{2&3&1&1\\   1&\bl\tcercle{1}}}
\end{split}
\end{equation}
giving then
\begin{gather*}
s^c_{(1;4)}=m_{(1;4)}+m_{(1;3,1)}+m_{(1;2,2)}+m_{(1;2,1,1)}+m_{(1;1,1,1,1)}
\\
\phantom{s^c_{(1;4)}=}{}
+m_{(2;3)}+m_{(2;2,1)}+m_{(2;1,1,1)}+m_{(3;2)}+m_{(3;1,1)}.
\end{gather*}
In the above expression, the ordering of the monomials follows that of the tableaux listed in~\eqref{tabex3}.
Here $m=1$ and~$\Omega_1^a$ is the number of boxes marked with~1 (these are identif\/ied with a~part in~$\Omega^a$ since
the circle is f\/illed with~1).
Then, the number of~2 gives the value of~$\Omega^s_1$, the number of~3 yields $\Omega^s_2$ and so on.
\end{example}

\begin{example}
Here is a~more complicated example, choosing $I_2=\{1,2\}$:
\[
\Lambda=(3,0;4,1)
\qquad
\tableau[scY]{&&\cdot&\cdot\\1&1&1&\bl\tcercle{1}\\ \\\bl\tcercle{2}}.
\]
Since $c_{2,1}=2$, there can be at most two~2 appearing in positions indicated with dotted boxes.
There is thus three possible f\/illings with fermionic numbers:
\[
\Omega^a=(3,0):
\quad
{\tableau[scY]{&&&\\1&1&1&\bl\tcercle{1}\\ \\\bl\tcercle{2}}}
\qquad
\Omega^a=(3,1):
\quad
{\tableau[scY]{&&&2\\1&1&1&\bl\tcercle{1}\\ \\\bl\tcercle{2}}}
\qquad
\Omega^a=(3,2):
\quad
{\tableau[scY]{&&2&2\\1&1&1&\bl\tcercle{1}\\ \\\bl\tcercle{2}}}
\]
with corresponding reduced diagrams $(4,1)$, $(3,1)$ and $(2,1)$.
We thus read
\begin{gather}
s^c_{(3,0;4,1)}=\sum\limits_{\Omega^s\leq (4,1)} K_{(4,1),\Omega^s} m_{(3,0;\Omega^s)}
+\sum\limits_{\Omega^s\leq(3,1)} K_{(3,1),\Omega^s} m_{(3,1;\Omega^s)}
\nonumber
\\
\phantom{s^c_{(3,0;4,1)}=}{}
+\sum\limits_{\Omega^s\leq (2,1)} K_{(2,1),\Omega^s} m_{(3,2;\Omega^s)},
\label{EX}
\end{gather}
so that
\begin{gather*}
s^c_{(3,0;4,1)}=m_{(3,0;4,1)}+m_{(3,0;3,2)}+2 m_{(3,0;3,1,1)}+2 m_{(3,0;2,2,1)}+3m_{(3,0;2,1,1,1)}
\\
\phantom{s^c_{(3,0;4,1)}=}{}
+4m_{(3,0;1,1,1,1,1)}
+m_{(3,1;3,1)}+m_{(3,1;2,2)}+2 m_{(3,1;2,1,1)}+3 m_{(3,1;1,1,1,1)}\\
\phantom{s^c_{(3,0;4,1)}=}{}
+m_{(3,2;2,1)}+2 m_{(3,2;1,1,1)}.
\end{gather*}
\end{example}

We now end this section with the announced conjecture that identif\/ies the Schur superpolynomials def\/ined
combinatorially, namely $s^c_\Lambda$, with those obtained from Macdonald superpolynomials for $q=t=0$, denoted~$s_\Lambda$ {(cf.~\eqref{2ss})}.

\begin{conjecture}
\label{ssc}
{We have $ s_\Lambda^c(x,\theta)=s_\Lambda(x,\theta)$.
Equivalently, $\bar{K}^c_{\Lambda \Omega}=\bar{K}_{\Lambda \Omega}$.}
\end{conjecture}

This conjecture has been extensively tested.
In particular, all the examples presented in this section agree with the statement of the conjecture.

\section{The Pieri rule}\label{Pieri}

We now present a~conjectural version of the   Pieri rule for the Schur superpolynomials $s_\Lambda$ def\/ined by~\eqref{2ss}.
In order to formulate the rule, we must recall the notion of horizontal and vertical~$k$-strips: a~horizontal (resp.\
vertical)~$k$-strip is a~skew diagram that has at most one square in each column (resp.\ row)~\cite{Mac}.
In the following, we need an extension that includes circles.

\begin{definition}[\cercle{$k$}-strips]
A~horizontal \cercle{$k$}-strip is a~horizontal~$k$-strip augmented by a~circle in its upper-right position; straightened
horizontally, it represents the diagram of the superpartition~$(k;)$.
A~vertical \cercle{$k$}-strip is a~vertical $k$-strip
augmented by a~circle in its lower-left position; straightened
vertically, it represents the diagram of the superpartition~$(0;1^k)$.
\end{definition}

Here are some examples of strips made of boxes marked by $\star$:
\begin{equation}
\label{ex0}
\tableau[scY]{&&&\bl\tcercle{$\star$}\\&&\star \\&\star&\bl\tcercle{} \\\star \\ \bl\tcercle{}\\ }
\qquad
\tableau[scY]{&&\star&\bl\tcercle{}\\&&\star \\&\star&\bl\tcercle{} \\\star \\ \bl\tcercle{$\star$}\\ }
\qquad
\tableau[scY]{&&&\bl\tcercle{}\\&&\star \\&\star&\bl\tcercle{} \\\star \\ \bl\tcercle{}\\ }
\end{equation}
representing respectively, a~horizontal \cercle{$3$}-strip, a~vertical \cercle{$4$}-strip and a~3-strip that is both horizontal and vertical.
The Pieri rule relies on a~specif\/ic rule, spelled out in the following, for the multiplication of a~row or a~column diagram with a generic diagram.

\begin{arule}[Multiplication of a~row/column with a~diagram~$\Lambda$]\label{StripInsert}\qquad
\begin{enumerate}\itemsep=0pt
\item[I.] Row {multiplication}: the squares and the circle (if the row is fermionic) that are added to the
diagram~$\Lambda$ must form a~horizontal strip, in addition to generate an admissible resulting diagram (i.e., rows are
weakly decreasing and there can be at most one circle per row and column).
Moreover, when inserting the squares of a~row into the diagram of~$\Lambda$, the circles of~$\Lambda$ can be displaced
subject to the following restrictions:
\begin{enumerate}\itemsep=0pt
\item[(i)] a~circle in the f\/irst row can be moved horizontally without restrictions;
\item[(ii)] a~circle not in the f\/irst row can be moved horizontally as long as there is a~square in the row just above
it in the original diagram~$\Lambda$ (i.e., the circle in row~$i$ can be displaced by at most
$\Lambda_{i-1}^*-\Lambda_i^*-1$ columns);
\item[(iii)] a~circle can be displaced vertically in the same column by at most one row.
\end{enumerate}
\item[II.] Column multiplication: interchange `row' and `column', `horizontal' and vertical', `above' and `at the
left' in I.
\end{enumerate}
\end{arule}

\begin{definition}[Pieri diagrams]
Let~$\Lambda$ and~$\Gamma$ be two superpartitions, with~$\Gamma$ {a row or a~column diagram}
(bosonic or fermionic).
We denote by~$\Lambda \otimes \Gamma$ the set of all admissibles diagrams, called Pieri diagrams, obtained by the
multiplication of~$\Gamma$ with the diagram of~$\Lambda$ using Rule~\ref{StripInsert}.
\end{definition}

Note that a~strip of either type needs not to be located completely on the exterior (or right) boundary of the larger diagram.
However, when the circles are erased, the strip is indeed at the exterior frontier of the diagram, which is clear from the examples~\eqref{ex0}.

Now, consider for example the multiplication of a~bosonic row of length~3 with the diagram of $\Lambda=(2,0;1)$, operation {being} denoted by
\[
\tableau[scY]{&&\bl\tcercle{}\\ \\ \bl\tcercle{}\\ }
\quad
\otimes
\quad
{\tableau[scY]{1&1&1}}\; ,
\]
where the boxes of the diagram $(;3)$ are marked by~1.
Using Rule~\ref{StripInsert}, the resulting Pieri diagrams (or tableaux, the diagrams being partially f\/illed) are
\begin{equation} \label{ex1}
\begin{split}
& {{\tableau[scY]{&&1&1&1&\bl\tcercle{}\\ \\ \bl\tcercle{}\\ }},
\qquad {\tableau[scY]{&&1&1&\bl\tcercle{}\\ &1\\ \bl\tcercle{}\\ }},
\qquad {\tableau[scY]{&&1&1&\bl\tcercle{}\\ \\1\\ \bl\tcercle{}\\ }},
\qquad {\tableau[scY]{&&1&\bl\tcercle{}\\ &1\\1\\ \bl\tcercle{}\\ }}},\\
&
{{\tableau[scY]{&&1\\&1&\bl\tcercle{}\\1 \\ \bl\tcercle{}\\ }},
\qquad {\tableau[scY]{&&1&1\\ &1&\bl\tcercle{}\\ \bl\tcercle{}\\ }}}.
\end{split}
\end{equation}
If, instead, we multiply a~fermionic row, we obtain
\[
{\tableau[scY]{&&\bl\tcercle{}\\ \\ \bl\tcercle{}\\ }} \quad \otimes \quad  {\tableau[scY]{1&1&1&\bl\tcercle{1} }}
\quad : \quad
 {\tableau[scY]{&&1&1&\bl\tcercle{1}\\&1&\bl\tcercle{} \\ \bl\tcercle{}\\ }},
\qquad {\tableau[scY]{&&1&\bl\tcercle{1}\\ &1&\bl\tcercle{}\\1\\ \bl\tcercle{}\\ }}.
\]

A similar example illustrating the multiplication of a~bosonic column, namely $(;1,1)$ (whose squares are marked~1 and~2)
into $\Lambda=(2,0;1)$, yields
\begin{equation}
\label{ex2f}
\tableau[scY]{&&\bl\tcercle{}\\ \\ \bl\tcercle{}\\ } \quad\!\!\! \otimes \quad\!  {\tableau[scY]{1 \\ 2  }}  \!\quad :
\quad
{\tableau[scY]{&&\bl\tcercle{}\\ \\1\\2 \\ \bl\tcercle{}\\ }},
\!\quad {\tableau[scY]{&&1&\bl\tcercle{}\\ \\2 \\ \bl\tcercle{}\\ }},
\!\quad {\tableau[scY]{&&\bl\tcercle{}\\ &1\\2 \\ \bl\tcercle{}\\ }},
\!\quad {\tableau[scY]{&&1&\bl\tcercle{}\\ &2 \\ \bl\tcercle{}\\ }},
\!\quad {\tableau[scY]{&&\bl\tcercle{}\\&1 \\2 & \bl\tcercle{}\\ }},
\end{equation}
while turning the column into a~fermionic one by the addition of a~circle marked 3, leads to a~single conf\/iguration,
namely
\[
{{\tableau[scY]{&&\bl\tcercle{}\\ \\ \bl\tcercle{}\\ }} \quad \otimes \quad  {\tableau[scY]{1 \\ 2  \\ \bl\tcercle{3} }}}
\quad : \quad
 {\tableau[scY]{&&\bl\tcercle{}\\&1 \\2 & \bl\tcercle{} \\\bl\tcercle{3}\\ }}.
\]

We are now in position to formulate the conjectural form of the Pieri rules.

\begin{conjecture}[Pieri formulas]
We have
\begin{gather}
\label{eqPierirow}
s_\Lambda s_{(; r)}=\sum\limits_{\Omega \in\Lambda\otimes (; r)}s_\Omega
\qquad
\text{and}
\qquad
s_\Lambda s_{(r; )}=\sum\limits_{\Omega \in\Lambda\otimes (r; )} (-1)^{\#\ell_\odot} s_\Omega
\end{gather}
and
\begin{gather*}
s_\Lambda s_{(;1^r)}=\sum\limits_{\Omega \in\Lambda\otimes(;1^r)} s_\Omega
\qquad
\text{and}
\qquad
s_\Lambda s_{(0; 1^r)}=\sum\limits_{\Omega \in\Lambda\otimes (0; 1^r)} (-1)^{\#\ell_\odot} s_\Omega.
\end{gather*}
The symbol $\#\ell_\odot$ stands for the number of circles in the diagram of~$\Omega$ that lie below the one which has
been added.
\end{conjecture}

Here are some examples illustrating these Pieri formulas.

\begin{example}
Consider the product $s_{(4,0;3)} s_{(3;)}$.
Using the multiplication Rule~\ref{StripInsert}, we f\/ind the following diagrammatic Schur superpolynomials
expansion:
\begin{equation}
\label{ex1t}
\begin{split}
{\tableau[scY]{&&&&\bl\tcercle{}\\&&\\\bl\tcercle{} \\ }}
\times {\tableau[scY]{ 1&1 &1& \bl\tcercle{1}\\ }}
=&
  \ {\tableau[scY]{&&&&1&1&\bl\tcercle{1}\\&&&1&\bl\tcercle{}\\\bl\tcercle{} \\ }}+
  {\tableau[scY]{&&&&1&\bl\tcercle{1}\\&&&1&\bl\tcercle{}\\1&\bl\tcercle{} \\ }}  \\ &
  {} +
    {\tableau[scY]{&&&&1&\bl\tcercle{1}\\&&&1&\bl\tcercle{}\\1\\\bl\tcercle{} \\ }}{-\;
      {\tableau[scY]{&&&&\bl\tcercle{}\\&&&\bl\tcercle{1}\\1&1&1\\ \bl\tcercle{} \\ }}}.
\end{split}
\end{equation}
Note that the last tableau appears with a~minus sign since there is one circle below the added one
(marked with~1)\footnote{The origin of the relative sign is clear from the algebraic point of view: since a~circle is associated
with a~factor~$\theta$ (the anticommuting variables), we see that, compared with the f\/irst three diagrams, the last one
is associated with a~dif\/ferent ordering of the f\/irst two~$\theta$ factors.
Upon reordering, this yields the minus sign.}.
Written in terms of the Schur superpolynomials, this reads
\begin{gather*}
s_{(4,0;3)} s_{(3;)}=s_{(6,4,0;)}+s_{(5,4,1;)}+s_{(5,4,0;1)}-s_{(4,3,0;3)}.
\end{gather*}

\noindent
{\bf Remark.}
Here one might wonder why no tableau appears where the bottom unmarked circle is moved
horizontally by two units (which is permitted by the length of the row just above) in~\eqref{ex1t}.
The only option would be
\[
{\tableau[scY]{&&&&1&\bl\tcercle{}\\&&&\bl\tcercle{1}\\1&1&\bl\tcercle{} \\}}.
\]
But the building strip is not a~horizontal ${3}$-strip: its upper-right component is a~square and not a~circle.
\end{example}

\begin{example}
Consider next the product $s_{(1;2,1)} s_{(0;1,1,1)}$ and f\/ill the column of the second diagram with numbers~1 to~4.
Using the Pieri formula, we have
\begin{equation}
\label{exfcpieri}
{\tableau[scY]{&\\&\bl\tcercle{}\\ \\ }}
\times {\tableau[scY]{ 1\\2 \\ 3\\ \bl\tcercle{4}\\ }}=
{\tableau[scY]{&\\&\bl\tcercle{}\\ \\ 1\\2\\3\\ \bl\tcercle{4}\\ }}+\;
{\tableau[scY]{&&1\\&\bl\tcercle{}\\ \\2\\3\\ \bl\tcercle{4}\\ }}+\;
{\tableau[scY]{&\\&1\\&\bl\tcercle{}\\2\\3\\ \bl\tcercle{4}\\ }}+\;
{\tableau[scY]{&&1\\&2\\ &\bl\tcercle{} \\3\\ \bl\tcercle{4}\\ }}+\;
{\tableau[scY]{&&1\\&2 &\bl\tcercle{} \\ \\3\\ \bl\tcercle{4}\\ }}+\;
{\tableau[scY]{&&1\\&2 &\bl\tcercle{} \\ &3\\ \bl\tcercle{4}\\ }}.
\end{equation}
The third and fourth tableaux illustrate the vertical motion of the circle.
The f\/ifth and sixth tableaux exemplify the allowed horizontal move by one column even if it exceeds the number of
squares of the previous row in the original tableau as long as this upper slot is occupied by a~square of the strip
(here the square marked~1).
Note that in all cases, the circle marked~4 is below the unmarked one: the factor $\#\ell_\odot$ is 0 in all diagrams,
so that we have
\begin{gather*}
s_{(1;2,1)} s_{(0;1^3)}=s_{(1,0;2,1^4)}+s_{(1,0;3,1^3)}+s_{(1,0;2^2,1^2)}+s_{(1,0;3,2,1)}+s_{(2,0;3,1^2)}+s_{(2,0;3,2)}.
\end{gather*}
{\bf Remark.}  Note that the following tableau
\[
{\tableau[scY]{&\\&1\\&2 \\ 3&\bl\tcercle{}\\ \bl\tcercle{4}\\}}
\]
is not allowed in the decomposition~\eqref{exfcpieri} since the top circle has moved vertically by two units.
This is forbidden because that gives the circle a~position exceeding the number of squares in the f\/irst column (cf.~point II(ii) of Rule~\ref{StripInsert}).
\end{example}

\begin{example}
Finally, from the examples~\eqref{ex1}, \eqref{ex2f}, we read of\/f:
\begin{gather*}
s_{(2,0;1)} s_{(;3)}=s_{(5,0;1)}+s_{(4,0;2)}+s_{(4,0;1,1)}+s_{(3,0;2,1)}+s_{(2,0;3,1)}+s_{(2,0;4)},
\\
s_{(2,0;1)} s_{(3;)}=s_{(4,2,0;0)}+s_{(3,2,0;1)},
\\
s_{(2,0;1)} s_{(;1^2)}=s_{(2,0;1^3)}+s_{(3,0;1^2)}+s_{(2,0;2,1)}+s_{(3,0;2)}+s_{(2,1;2)},
\\
 s_{(2,0;1)} s_{(0;1^2)}=s_{(2,1,0;2)}.
\end{gather*}
\end{example}

Let us compare the tableaux resulting from the successive applications of the row-version of the Pieri rule with the
super semi-standard tableaux described in Section~\ref{SST}.
Consider for instance a~complete f\/illing of the tableau at the left in~\eqref{exfixedpair}
\[
\tableau[scY]{3&3&3&5&2&2\\4&4&4&2&1  \\ 1&1&1 &\bl \tcercle{1} \\ 5&5 \\ \bl\tcercle{2}}
\]
for which the ordering is $1>2>5>4>3$.
Clearly, there is no way to remove successively horizontal strips composed of 1, and then of 2, 5, etc.
Therefore, although the Pieri rule builds up Schur superpolynomials, there is no relation between the Kostka numbers
$\bar K_{\Lambda\Omega}$ and this Pieri rule.
It turns out that the latter is directly related to the {\it dual} semi-standard tableaux enumerated by the Kostka
numbers $K_{\Lambda\Omega}$, namely, the expansion coef\/f\/icients of the dual Schur superpolynomials~$\bar s_\Lambda$ in
the monomial basis.
These dual tableaux are introduced in Section~\ref{dual}.

\section{The combinatorial dual Schur superpolynomials} \label{dual}

In this section, we introduce a~second family of combinatorial functions, $\bar{s}^c_\Lambda$, {def\/ined} in terms of
a~dual counterpart of the super semi-standard tableaux.
The main result here is the conjectural equivalence of these new functions with the~$\bar{s}_\Lambda$.

\subsection{Dual super semi-standard tableaux}

We thus f\/irst introduce the dual tableaux.
The qualitative {\it dual} refers to the way the fermionic numbers are ordered and the rules for their insertion in the tableau.
Rule~\ref{numcirc} still holds and the set~$I_m$ is def\/ined as before, as the ordered set of the labels in circles read from top to bottom.
But Rule~\ref{order} is replaced by its dual.

\begin{arule}[dual ordering in the set~$I$]\label{dorder}
The numbers in the set $I_m$ are {still} considered to be the~$m$ largest numbers of the set~$I$ but they are now
ordered as $i_m>i_{m-1}>\dots >i_1$ (and $i_1>j$ $\forall\, j\in I_m^c$).
In $I_m^c$, the ordering is the natural one.
\end{arule}

For example~\eqref{exfill}, this yields $5>1>3>14>13>\cdots>2$.

As already indicated, the rules for the f\/illing of the fermionic numbers are also modif\/ied.
At f\/irst, the numbers in the fermionic rows are no longer frozen.
In addition, the fermionic number~$i_k$, when appearing in a~bosonic row, is forbidden in the squares at the upper-right
of the circled~$i_k$; it can only appear in the dual region, at its lower-left.
This is made precise in the following.

\begin{arule}[fermionic numbers]\label{dfer}
The occurrence of the fermionic numbers in boxes must satisfy the following conditions:
\begin{enumerate}\itemsep=0pt
\item[a)] the number $i_k$ can only appear at the left of the column ending with circled $i_k$;
\item[b)] the fermionic number $i_{k-\ell}$, with $\ell\geq1$, must appear above or at the right of the circled $i_{k}$
at least~$\ell$ times;
\item[c)] counting the boxes from right to left and top to bottom, the number of boxes marked $i_k$ must always be
strictly greater than those marked $i_{k+1}$.
\end{enumerate}
\end{arule}

 For instance, the fermionic f\/illing (with $I_2=\{1,2\}$) of the shape $(5,3;1)$ and content $ (3,2;1^4)$ are
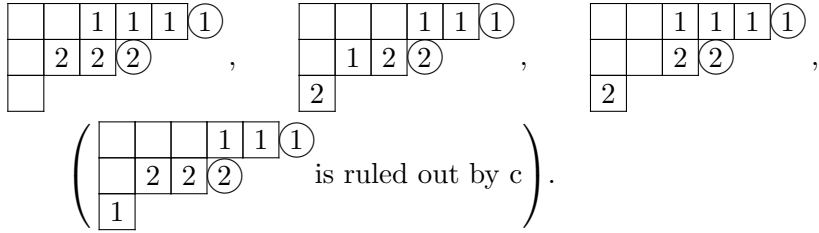
\begin{equation*}
\begin{split}
& {{\tableau[scY]{&&1&1&1&\bl\tcercle{1}\\&2&2&\bl\tcercle{2}\\ \\ }}} \, , \qquad
{{\tableau[scY]{&&&1&1&\bl\tcercle{1}\\&1&2&\bl\tcercle{2}\\ 2 \\ }}} \,,\qquad
{{\tableau[scY]{&&1&1&1&\bl\tcercle{1}\\&&2&\bl\tcercle{2}\\ 2\\ }}} \,,\qquad  \\
& \qquad
{\left(\;\tableau[scY]{&&&1&1&\bl\tcercle{1}\\&2&2&\bl\tcercle{2}\\ 1\\} \text{is ruled out  by {c}}\right)}.
\end{split}
\end{equation*}
To illustrate Rule~\ref{dfer}b, consider the diagram $(8,4,2,0;6,3,3,1,1)$ and the two fermionic f\/illings of content
$\Omega^a=(4,2,1,0)$:
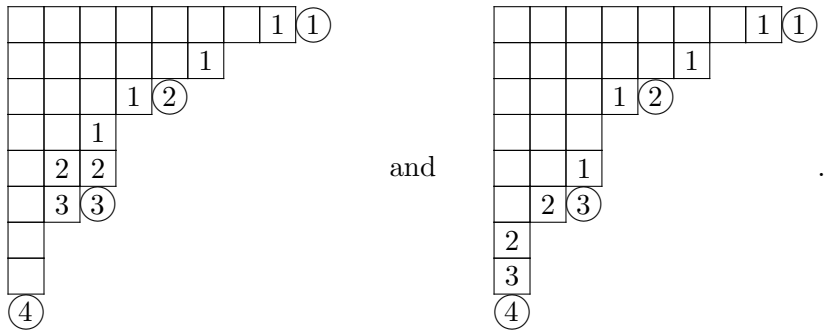
\[
\tableau[scY]{&&&&&&&1&\bl\tcercle{1}\\ &&&&&1\\&&&1&\bl\tcercle{2}\\ &&1\\&2&2\\&3&\bl\tcercle{3}\\ \\ \\\bl\tcercle{4}}\qquad
\text{and}\qquad
\tableau[scY]{&&&&&&&1&\bl\tcercle{1}\\ &&&&&1\\&&&1&\bl\tcercle{2}\\ &&\\&&1\\&2&\bl\tcercle{3}\\ 2\\3 \\\bl\tcercle{4}}.\qquad
\]
Let AR stands for `above or at the right'.
The conditions on the appearance of the number~1 are: once AR of~$2$, twice AR of~$3$, three times AR of~$4$.
Similarly, the number 2 must appear at least once AR of~$3$, and twice AR of~$4$, while the~3 must appear once AR of~$4$.
The tableau at the left satisf\/ies all the conditions but not the one at the right (the f\/irst condition on the number~2 is violated).

Finally, Rule~\ref{SSfil} needs to be trivially modif\/ied as follows.

\begin{arule}[dual semi-standard f\/illing]\label{DSSfil}
The boxes are f\/illed with numbers in~$I$ using the ordering def\/ined in Rule~\ref{dorder} and such that numbers in rows
are weakly increasing and strictly increasing in columns.
\end{arule}

\begin{definition}
A~tableau $\bar T^\circ$ satisfying Rules~\ref{numcirc},~\ref{dorder},~\ref{dfer} and~\ref{DSSfil} is called a~dual super semi-standard tableau.
\end{definition}

 For instance, the dual semi-standard tableaux with shape $\Lambda=(2,0;2,1)$ and $I_2=\{1,2\}$ are
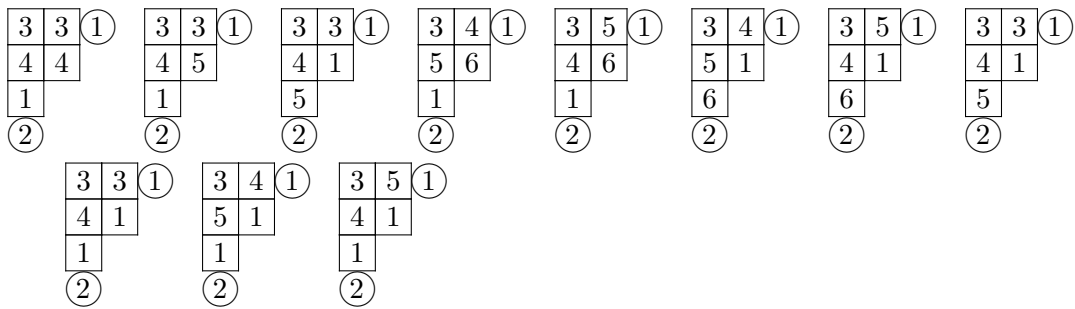
\begin{equation*}
\begin{split}
& {{\tableau[scY]{3&3&\bl\tcercle{1}\\4&4\\1 \\\bl\tcercle{2}}}} \quad
 {{\tableau[scY]{3&3&\bl\tcercle{1}\\4&5\\1\\\bl\tcercle{2}}}} \quad
 {{\tableau[scY]{3&3&\bl\tcercle{1}\\4&1\\5\\\bl\tcercle{2}}}} \quad
 {{\tableau[scY]{3&4&\bl\tcercle{1}\\5&6\\1 \\\bl\tcercle{2}}}} \quad
 {{\tableau[scY]{3&5&\bl\tcercle{1}\\4&6\\1 \\\bl\tcercle{2}}}} \quad
 {{\tableau[scY]{3&4&\bl\tcercle{1}\\5&1\\6 \\\bl\tcercle{2}}}} \quad
 {{\tableau[scY]{3&5&\bl\tcercle{1}\\4&1\\6 \\\bl\tcercle{2}}}} \quad
 {{\tableau[scY]{3&3&\bl\tcercle{1}\\4&1\\5 \\\bl\tcercle{2}}}} \quad \\
& \qquad
 {{\tableau[scY]{3&3&\bl\tcercle{1}\\4&1\\1 \\\bl\tcercle{2}}}} \quad
 {{\tableau[scY]{3&4&\bl\tcercle{1}\\5&1\\1 \\\bl\tcercle{2}}}} \quad
 {{\tableau[scY]{3&5&\bl\tcercle{1}\\4&1\\1 \\\bl\tcercle{2}}}} \quad
\end{split}
\end{equation*}
Notice that at least one box must have~1 (i.e., set $k=2$ and $\ell=1$ in Rule~\ref{dfer}b) and none can have~2.

\subsection{Combinatorial def\/inition of the dual Schur superpolynomials}

We are now in position to def\/ine the combinatorial dual Schur superpolynomials.
\begin{definition}
The combinatorial dual Schur superpolynomial $\bar{s}_\Lambda^c=\bar{s}_\Lambda^c(x,\theta)$ is def\/ined as
\begin{gather*}
\bar{s}^c_\Lambda:=\sum\limits_{\bar T^\circ\in \bar{\mathcal T}^\circ(\Lambda)} \zeta^{\bar T^\circ},
\end{gather*}
where $\bar{\mathcal T}^\circ(\Lambda)$ is the set of dual semi-standard tableaux of shape~$\Lambda$.
\end{definition}

For example, the superpolynomial $\bar{s}_{(1;2)}^c$ is obtained from the tableaux:
\begin{equation*}
\begin{split}
& {{\tableau[scY]{2&2 \\1& \bl\tcercle{1} }}} \; , \quad
{{\tableau[scY]{1&1 \\2& \bl\tcercle{2} }}} \; ,  \quad \ldots , \quad
{{\tableau[scY]{2&3 \\1& \bl\tcercle{1} }}} \; , \quad
{{\tableau[scY]{1&3 \\2& \bl\tcercle{2} }}} \; , \quad \cdots , \quad
{{\tableau[scY]{2&2 \\3& \bl\tcercle{1} }}} \; , \quad
{{\tableau[scY]{1&1 \\3& \bl\tcercle{2} }}} \; , \quad \ldots ,   \\
& \qquad
{{\tableau[scY]{2&3 \\4& \bl\tcercle{1} }}} \; , \quad
{{\tableau[scY]{2&4 \\3& \bl\tcercle{1} }}} \; , \quad
{{\tableau[scY]{1&3 \\4& \bl\tcercle{2} }}} \; , \quad
{{\tableau[scY]{1&4 \\3& \bl\tcercle{2} }}}\; , \quad \dots,
\end{split}
\end{equation*}
which reads
\begin{gather*}
\theta_1 x_1 x_2^2+\theta_2 x_1^2 x_2+\cdots+\theta_1 x_1 x_2 x_3+\theta_2 x_1 x_2 x_3+\cdots+\theta_1 x_2^2x_3
\\
\qquad
{}+\theta_2 x_1^2 x_3+\cdots+2 \theta_1 x_2 x_3 x_4+2 \theta_2 x_1 x_3 x_4+\cdots.
\end{gather*}

\begin{proposition}
The $\bar{s}_\Lambda^c$ are symmetric superpolynomials.
\end{proposition}

The proof follows that establishing the symmetric character of $s^c_\Lambda$ (cf.~Proposition~\ref{sisi}) and will
be omitted.

\begin{proposition}
We have
\begin{gather*}
\bar{s}_\Lambda^c=\sum\limits_{\Omega} K_{\Lambda \Omega}^c m_\Omega,
\end{gather*}
where $K^c_{\Lambda \Omega}$ is the number of elements of $\bar{\mathcal T}^\circ(\Lambda)$ with content~$\Omega$ and
it satisfies
\begin{gather*}
{K}^c_{\Lambda\Omega}=
\begin{cases}
0&\text{if}\quad \Omega \not \leq \Lambda,
\\
1&\text{if}\quad \Omega=\Lambda.
\end{cases}
\end{gather*}
\end{proposition}

This proof will also be omitted.

\begin{example}
\label{EEX}
Let us evaluate the number $K^c_{(8,4;1,1), (6,3;1^5)}$.
The allowed fermionic-number f\/illings, with $I_2=\{1,2\}$, are
\begin{equation*}
 \tableau[scY]{&&1&1&1&1&1&1& \bl \tcercle{1} \\&2 &2&2& \bl\tcercle{2}  \\ \\ \\}, \qquad
\tableau[scY]{&&&&1&1&1&1& \bl \tcercle{1} \\&1&2&2& \bl\tcercle{2}  \\ 1\\2 \\}, \qquad
\tableau[scY]{&&1&1&1&1&1&1& \bl \tcercle{1} \\& &2&2& \bl\tcercle{2}  \\ \\ 2\\},
\end{equation*}
\begin{equation*}
 \tableau[scY]{&&&1&1&1&1&1& \bl \tcercle{1} \\& &2&2& \bl\tcercle{2}  \\ 1\\2 \\}, \qquad
\tableau[scY]{&&&1&1&1&1&1& \bl \tcercle{1} \\&1 &2&2& \bl\tcercle{2}  \\ \\2 \\}, \qquad
\tableau[scY]{&&&1&1&1&1&1& \bl \tcercle{1} \\&2 &2&2& \bl\tcercle{2}  \\ \\1 \\}.
\end{equation*}
The multiplicity of each tableau is equal to the multiplicity of the corresponding reduced standard (i.e., with all
entries being distinct) tableaux.
This gives multiplicity~4 for the f\/irst two tableaux, 5 to the following two and~6 for the remaining ones   which gives
a~total of~30, i.e., $K^c_{(8,4;1,1), (6,3;1^5)}=30$.
\end{example}

\begin{conjecture}
\label{bscebs}
{We have $\bar{s}^c_\Lambda=\bar{s}_\Lambda$.
Equivalently, $K_{\Lambda \Omega}^c=K_{\Lambda \Omega}$.}
\end{conjecture}

Although there is no relation between the two versions of the Schur superpolynomials, the next section {reveals} a~striking {indirect} connection.

\section{Relating the Pieri rule to dual semi-standard tableaux}\label{Section6}

As indicated at the end of Section~\ref{Pieri}, the Pieri tableaux obtained by successive row multiplications do not correspond to
super semi-standard tableaux.
They are rather related to their dual versions.
This is a~consequence of the nontrivial duality property of the Schur superpolynomials.
Let $\langle \cdot, \cdot \rangle$ be the scalar product~\eqref{psqt} with $q=t=1$:
\begin{gather*}
\langle p_\Lambda, p_\Omega \rangle=(-1)^{\binom{m}{2}} \delta_{\Lambda \Omega}z_{\Lambda^s}.
\end{gather*}
This scalar product is equivalent to
\begin{gather*}
\langle s_\Lambda^*, s_\Omega \rangle=\delta_{\Lambda \Omega},
\end{gather*}
where $s_\Lambda^*:=(-1)^{\binom{m}{2}}\omega(\bar{s}_{\Lambda'})$ and~$\omega$ being the involution def\/ined
as~\cite[Corollary~29]{BDLM2}
\begin{gather*}
\omega(p_r)=(-1)^{r-1}p_r,
\qquad
\omega(\tilde{p}_{r-1})=(-1)^{r-1}\tilde{p}_{r-1}.
\end{gather*}
Now, if we def\/ine a~new basis $H_\Lambda$ given by $H_\Lambda:=\tilde{p}_{\Lambda^a} h_{\Lambda^s}$, we observe that
\begin{gather*}
\langle s^*_\Lambda, H_\Omega \rangle=K_{\Lambda \Omega},
\end{gather*}
or, equivalently
\begin{gather}
\label{HvsS}
H_\Lambda=\sum\limits_{\Omega \geq \Lambda}K_{\Omega \Lambda}s_\Omega.
\end{gather}
Setting $\Lambda=(n; )$ in the previous relation implies that $\tilde{p}_n=s_{(n;)}$.
Indeed, $(n; )$ is the largest superpartition with degree $(n|1)$ so that there is a~single contributing term in the
sum, and its coef\/f\/icient is $K_{\Lambda \Lambda}=1$.
The equation~\eqref{HvsS} can thus be written as
\begin{gather*}
s_{(\Lambda_1; )} \cdots s_{(\Lambda_m; )} s_{(; \Lambda_{m+1})} \cdots s_{(; \Lambda_N)}=\sum\limits_{\Omega \geq\Lambda}K_{\Omega \Lambda}s_\Omega.
\end{gather*}
This implies that by using the Pieri rules for multiplication of rows corresponding to the parts of the
superpartition~$\Lambda$, the number of admissible Pieri diagrams (erasing the f\/illing) equal to~$\Omega$ must be
identical to the number of dual semi-standard tableaux of shape~$\Omega$ and content~$\Lambda$, which is precisely the
number $K_{\Omega \Lambda}^c$ from Conjecture~\ref{bscebs}.
Note that this connection ensures the positivity of the Pieri rule, which is not manifest from~\eqref{eqPierirow} (but
see the remark below).

In order to make this connection more precise, let $P^\circ(\Lambda)$ be the set of all admissible Pieri tableaux
resulting from successive row multiplications, the rows corresponding to the parts of the superpartition~$\Lambda$
\begin{gather}
\label{srowmulord1}
(; \Lambda_{m+1}) \otimes \cdots \otimes (; \Lambda_\ell) \otimes (\Lambda_1; ) \otimes \cdots \otimes (\Lambda_m; ).
\end{gather}
In this multiple product, the order is taken to be from left to right.
We thus start row multiplications in the bosonic sector and end up with the multiplication of the fermionic rows.

As an illustrative example of such computations, consider $P^\circ((0;1^3))$.
Let the row associated with $\Lambda_i$ be marked with number~$i$.
The f\/irst step, the bosonic-row multiplications, yields
\begin{equation*}
\left(\,\tableau[scY]{2 \\}\otimes\tableau[scY]{3\\}\,\right)\otimes {{\tableau[scY]{4\ }}}={{\left(
\,
{{\tableau[scY]{2\\3 \\ }}}\,,
\;
{{\tableau[scY]{2&3 \\ }}}
\,
\right)}} \otimes {{\tableau[scY]{4 \\ }}}={{\tableau[scY]{2&4 \\3 \\ }}}
\,
,
\qquad
\tableau[scY]{2\\3\\4\\ }
\,
,
\qquad
\tableau[scY]{2&3&4\\ }
\,
,
\qquad
\tableau[scY]{2&3\\4\\ }.
\end{equation*}
Next, the multiplication with the sole fermionic row $(0; )$ gives
\begin{equation}
\label{EX1}
\begin{split}
&{{\tableau[scY]{2 & 4 & \bl\tcercle{1} \\ 3 \\  }}}, \qquad
{{\tableau[scY]{2 & 4  \\ 3 & \bl\tcercle{1} \\  }}} \;, \qquad
{{\tableau[scY]{2 & 4  \\ 3  \\  \bl\tcercle{1} }}} , \qquad
{{\tableau[scY]{2 & \bl\tcercle{1} \\ 3 \\ 4 \\  } }} , \qquad
{{\tableau[scY]{2 \\ 3 \\ 4 \\  \bl\tcercle{1}    }}}  \; ,   \\
&
{{\tableau[scY]{2&3 &4 & \bl\tcercle{1}   \\ }}} \; , \qquad
{{\tableau[scY]{2&3 &4   \\ \bl\tcercle{1} }}} \; , \qquad
{{\tableau[scY]{2&3 & \bl\tcercle{1}   \\ 4\\ }}} , \qquad
{{\tableau[scY]{2&3    \\ 4 & \bl\tcercle{1}\\ }}} \;, \qquad
{{\tableau[scY]{2&3    \\ 4 \\  \bl\tcercle{1}  }}}. \end{split}
\end{equation}

This relationship entails another combinatorial def\/inition of the Kostka numbers $K_{\Lambda\Omega}$.
\begin{conjecture}
We have
\begin{gather*}
K_{\Lambda \Omega}=\sum\limits_{T^\circ\in P^\circ(\Omega, \Lambda)} (-1)^{\inv(T^\circ)},
\end{gather*}
where $P^\circ(\Lambda, \Omega)$ is the set of tableaux of $P^\circ(\Lambda)$ that have shape~$\Omega$ and
$\inv(T^\circ)$ is the minimal number of permutations needed to reorder the fermionic set in $T^\circ$ from top to bottom in increasing order.
\end{conjecture}

 By Conjecture~\ref{bscebs}, this corresponds to the number of dual super semi-standard tableaux, which are
enumerated by $K^c_{\Lambda\Omega}$.
In the above example (cf.~\eqref{EX1}), we have displayed the dual super semi-standard tableaux with content $(0;1^3)$
for all superpartitions of degree $(3|1)$.
As another example, consider $P^\circ((2,1;1^3), (3,2;1))$; it is given by the tableaux:
\[
\underset{1}{\tableau[scY]{&1& 1& \bl\tcercle{1} \\&2& \bl\tcercle{2} \\ \\ } }, \quad
\underset{2}{\tableau[scY]{&& 1& \bl\tcercle{1} \\ &1& \bl\tcercle{2} \\2 \\  }} , \quad
\underset{-2}{\tableau[scY]{&& 2& \bl\tcercle{2} \\&1& \bl\tcercle{1} \\1 \\ } }, \quad
\underset{2}{\tableau[scY]{&&1 & \bl\tcercle{1} \\&2& \bl\tcercle{2} \\ 1\\ }  },\quad
\underset{1}{\tableau[scY]{&& & \bl\tcercle{1} \\1&1& \bl\tcercle{2} \\ 2\\  }} , \quad
\underset{-1}{\tableau[scY]{&& & \bl\tcercle{2} \\1&1& \bl\tcercle{1} \\2 \\  } },
\]
where the multiplicity below each tableau corresponds to the number of standard f\/illings of the empty boxes with the
omitted numbers $3$, $4$, $5$ times $(-1)^{\inv(T^\circ)}$.
Summing all contributions gives a~total of $3$ which is precisely the value of $K_{(3,2;1),(2,1;1^3)}$.

\medskip

\noindent
{\bf Remark.} The ordering used in~\eqref{srowmulord1} is motivated by the ordering f\/ixed by Rule~\ref{dorder}.
However, we can construct the Pieri tableaux using a~dif\/ferent multiplication order, for instance, beginning with the multiplication
of fermionic rows and taking the usual number ordering: $1<2<\cdots <m<m+1<\cdots$ (and distinguishing this set of tableaux with a~tilde):
\begin{gather*}
\tilde{P}^\circ(\Lambda) :
\
(\Lambda_1; ) \otimes \cdots \otimes (\Lambda_m; ) \otimes (; \Lambda_{m+1}) \otimes \cdots \otimes (; \Lambda_\ell)
\end{gather*}
(multiplying as usual from left to right).
One advantage of this procedure is that the product of the~$m$ fermionic rows generate a~single tableau with a~positive
sign.
Because the remaining multiplication of bosonic rows is manifestly positive, this ensures the positivity of the
super-extension of the Pieri rules.
This gives another combinatorial def\/inition for the dual super semi-standard tableaux: $K_{\Omega \Lambda}$ is given by
the cardinality of the set $\tilde{P}^\circ(\Lambda,\Omega)$.
For the above example, $\Lambda=(2,1;1^3)$ and $\Omega=(3,2;1)$, after the f\/irst step one has
\[
\tableau[scY]{1 &1 & \bl\tcercle{1} \\  2& \bl\tcercle{2}} \otimes \tableau[scY]{3 \\ } \otimes \tableau[scY]{4\\ } \otimes \tableau[scY]{5\\ }\,,
\]
which gives for $\tilde{P}^\circ((2,1;1^3),(3,2;1))$:
\[
{\tableau[scY]{1&1& 3& \bl\tcercle{1} \\2&4& \bl\tcercle{2} \\5 \\ } }, \qquad
{\tableau[scY]{1&1& 3& \bl\tcercle{1} \\ 2&5& \bl\tcercle{2} \\4 \\  }} , \qquad
{\tableau[scY]{1&1& 4& \bl\tcercle{{1}} \\2&5& \bl\tcercle{2} \\3 \\ }} .
\]
This ordering could be used to formulate new rules for the construction of dual super semi-standard tableaux (rules
which we have not found however).
But we then lose the connection with the usual Kostka numbers that enumerate the f\/illing of the reduced tableaux
obtained once the fermionic numbers have been introduced (cf.~Example~\ref{EEX} and~\eqref{EX} for a~similar result
in the non-dual context), which connection is a~great computational advantage.

\appendix

\section{Superpartitions and symmetric superpolynomials} 

In this f\/irst appendix, we {summarize the basic} notions and def\/initions {pertaining to} symmetric superpolynomials.

\subsection{Superpartitions}\label{Spart}

Superpartitions are generalizations of regular partitions and are the combinatorial objets used to label symmetric superpolynomials.
We f\/irst give the following def\/inition.

\begin{definition}[superpartition: the $(\Lambda^a;\Lambda^s)$ description]\label{SP1}
A~superpartition~$\Lambda$ is a~pair of partitions~\cite{DLM1}
\begin{gather*}
\Lambda=(\Lambda^{a};\Lambda^{s})=(\Lambda_1,\ldots,\Lambda_m;\Lambda_{m+1},\ldots,\Lambda_\ell),
\end{gather*}
such that
\begin{gather*}
\Lambda_1>\cdots>\Lambda_m\geq0
\qquad
\text{and}
\qquad
\Lambda_{m+1}\geq \Lambda_{m+2} \geq \cdots \geq \Lambda_\ell > 0.
\end{gather*}
We stress that $\Lambda^a$ has distinct parts and the last part is allowed to be 0.
The number~$m$ is called the fermionic degree of~$\Lambda$ and $n=|\Lambda|=\sum\limits_i\Lambda_i$ is the bosonic
degree.
Such superpartition~$\Lambda$ is said to be of degree $(n|m)$, {which is denoted as $\Lambda\vdash(n|m)$.}
\end{definition}

The diagrammatic representation of superpartitions is very similar to the usual Young diagrams representing
partitions. By removing the semi-coma and reordering the parts of~$\Lambda$, we obtain an ordinary partition that we
denote $\Lambda^*$.
The diagram of~$\Lambda$ is that of $\Lambda^*$ with circles added to the rows corresponding to the parts of $\Lambda^a$
and ordered in length as if a~circle was a~half-box~\cite{DLMclass}.
Here is an example of a~superpartition with degree $(27|5)$:
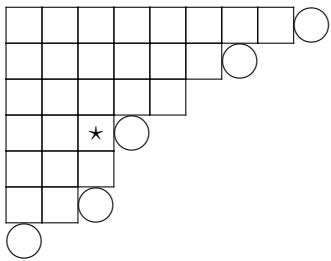
\[
\tableau[scY]{&&&&&&&&\bl\tcercle{}\\&&&&&&\bl\tcercle{}\\&&&&\\&& \star& \bl\tcercle{}\\&&\\&&\bl\tcercle{}\\\bl\tcercle{}}
\]
which corresponds to $\Lambda=(8,6,3,2,0;5,3)$.
Each box and circle in the diagram of~$\Lambda$ can be identif\/ied by its position $s=(i,j)$, where~$i$ denotes the row,
numbered from top to bottom, and~$j$ denotes the column, numbered from left to right.
A~row or column ending with a~circle is dubbed {\it fermionic}.
Other rows and columns are said to be {\it bosonic}.
In the above diagram, the box with a~$\star$ has position $s=(4,3)$; it belongs to the third fermionic row and the
second fermionic column.
With this diagrammatic representation, it is simple to def\/ine the conjugate operation.

\begin{definition}[conjugate superpartition]
The diagram of the superpartition $\Lambda'$ conjugate of~$\Lambda$ is obtained by interchanging the rows and columns of
the diagram~$\Lambda$.
\end{definition}

For example,
\[
\left(\;{\tableau[scY]{&&\bl\tcercle{}\\\bl\tcercle{} }}\right)'={\tableau[scY]{& \bl\tcercle{}\\ \\\bl\tcercle{} }}
\quad
\implies
\quad (20;)'=(1,0;1).
\]

\begin{definition}[superpartition: the $\Lambda^*$, $\Lambda^{\circledast}$ description]\label{SP3}
A~superpartition~$\Lambda$ of degree $(n|m)$ is described by a~pair of~2 partitions $\Lambda^*$, $\Lambda^\circledast$,
\begin{gather*}
\Lambda
\quad
\Leftrightarrow
\quad
\Lambda^*, \Lambda^\circledast
\end{gather*}
such that $|\Lambda^*|=|\Lambda^\circledast|-m=n$ and $\Lambda^\circledast/\Lambda^*$ {is a~skew diagram that is both}
a~horizontal and vertical~$m$-strip.
Diagrammatically, $\Lambda^*$ is obtained by removing all circles in~$\Lambda$ and $\Lambda^\circledast$ is obtained
by replacing all the circles by boxes.
\end{definition}

Clearly, both def\/initions~\ref{SP1} and~\ref{SP3} completely characterizes~$\Lambda$.
For example, with $\Lambda=(4,3,0;4)$:
\[
\Lambda={\tableau[scY]{&&&&\bl\tcercle{}\\&&&\\&&&\bl\tcercle{}\\\bl\tcercle{} \\ }}
\quad \Longleftrightarrow \quad \Lambda^\circledast={\tableau[scY]{&&&&\\&&& \\&&&\\  \\ }}, \quad  \Lambda^*={\tableau[scY]{&&&\\&&&\\ &&\\ \bl\\ }}\,.
\]

We now introduce the version of the dominance ordering that applies to superpartitions; it relies on the
$\Lambda^*$, $\Lambda^{\circledast}$ representation.

\begin{definition}[dominance order]\label{dom}
We say that~\cite{DLMeva}:
\begin{gather*}
\Lambda \geq \Omega
\quad
\iff
\quad
\Lambda^* \geq \Omega^*
\qquad
\text{and}
\qquad
\Lambda^{\circledast} \geq {\Omega^{\circledast}},
\end{gather*}
where, recall that, the order on partitions is the usual dominance ordering:
\begin{gather*}
\lambda \geq \mu
\quad
\iff
\quad
{|\lambda|=|\mu|}
\qquad
\text{and}
\qquad
\lambda_1+\cdots+\lambda_k \geq \mu_1+\cdots+\mu_k
\qquad
\forall\, k.
\end{gather*}
\end{definition}

Pictorially, $\Lambda {\geq}\Omega$ if~$\Omega$ can be obtained from~$\Lambda$ successively by moving down a~box or a~circle.
For example,
\[
\tableau[scY]{&& \bl\tcercle{}  \\ &} \geq \tableau[scY]{& \\ \\ \\ \bl\tcercle{}}
\qquad
\text{but}
\qquad
\tableau[scY]{&& \bl\tcercle{}  \\ &} \; \not \geq  \; \tableau[scY]{&&  \\ \\ \bl\tcercle{} }
\quad
\text{and}
\quad
\tableau[scY]{&&  \\ \\ \bl\tcercle{} }   \; \not \geq  \;\tableau[scY]{&& \bl\tcercle{}  \\ &}.
\]
In the latter two cases, the superpartitions are non-comparable.

\subsection{Symmetric superpolynomials}

Superpolynomials are polynomials in the usual commuting~$N$ variables $x_1,\ldots,x_N$ and the~$N$ anticommuting
variables $\theta_1,\ldots,\theta_N$.
Symmetric superpolynomials are invariant with respect to the interchange of $(x_i,\theta_i)\leftrightarrow
(x_j,\theta_j)$ for any $i$, $j$~\cite{DLM1}.

The space of symmetric superpolynomials, denoted $\mathcal{R}^{S_N}=F[x,\theta]^{S_N}$ where~$F$ is some f\/ield, is
naturally graded:
\begin{gather*}
\mathcal{R}^{S_N}=\bigoplus_{n,m} \mathcal{R}^{S_N}_{(n|m)},
\end{gather*}
where $\mathcal{R}^{S_N}_{(n|m)}$ is the space of homogeneous symmetric superpolynomials of degree~$n$
in the~$x$ variables and degree~$m$ in the~$\theta$ variables.
Bases of $\mathcal{R}^{S_N}_{(n|m)}$ are labelled by superpartitions of degree $(n|m)$.
We now present the superpolynomial version of the classical bases~\cite{DLMclass}.

\begin{definition}[super-monomials]\label{SM}
The super-version of the monomial function~\eqref{mono} reads:
\begin{gather*}
m_\Lambda(x,\theta)=\theta_{1} \cdots\theta_{m} x_1^{\Lambda_1} \cdots x_N^{\Lambda_N}+\text{distinct permutations of
$(x_i,\theta_i)\leftrightarrow (x_j,\theta_j)$}.
\end{gather*}
(Here and elsewhere, it is understood that $\Lambda_{\ell+1}=\dots=\Lambda_N=0$.)
\end{definition}

Here is an example, for $N=4$:
\begin{gather*}
m_{(1,0;1,1)}(x;\theta)=\theta_1\theta_2(x_{1}-x_2)x_3x_4+\theta_1\theta_3(x_{1}-x_3)x_2x_4+\theta_1\theta_4(x_{1}-x_4)x_2x_3
\\
\phantom{m_{(1,0;1,1)}(x;\theta)=}{}
+\theta_2\theta_3(x_{2}-x_3)x_1x_4+\theta_2\theta_4(x_{2}-x_4)x_1x_3+\theta_3\theta_4(x_{3}-x_4)x_1x_2.
\end{gather*}
Another example illustrating the fact that $\Lambda_m$ is allowed to be $0$ is $m_{(0;3)}$, written here for $N=2$:
\begin{gather*}
m_{(0;3)}(x;\theta)=\theta_1x_2^3+\theta_2x_1^3.
\end{gather*}

\begin{definition}[multiplicative bases]\label{MulSB}
For a~superpartition~$\Lambda$, we have the following mul\-tiplicative basis
\begin{gather*}
f_\Lambda=\tilde{f}_{\Lambda_1} \cdots \tilde{f}_{\Lambda_m} f_{\Lambda_{m+1}} \cdots f_{\Lambda_\ell},
\end{gather*}
where $f_\Lambda$ stands for:

1) the super-power-sums:
\begin{gather}
\label{Spower}
\tilde{p}_r=\sum\limits_i\theta_ix_i^r
\qquad
\text{and}
\qquad
p_s=\sum\limits_ix_i^s,
\end{gather}

2) the elementary superpolynomials:
\begin{gather*}
\tilde{e}_r=m_{(0;1^r)}
\qquad
\text{and}
\qquad
e_s=m_{(1^s)},
\end{gather*}

3) the completely homogeneous symmetric superpolynomials:
\begin{gather*}
\tilde{h}_r=\sum\limits_{\Lambda\vdash(r|1)}(\Lambda_1+1)m_{\Lambda}
\qquad
\text{and}
\qquad
h_s=\sum\limits_{\lambda\vdash(s|0)}{m_{\lambda}}
\end{gather*}
In the three cases, we have $r\geq 0$ and $s\geq 1$.
\end{definition}

\section{Bilinear identities in superspace}\label{Bil}

We end this article with the presentation of a~series of bilinear identities for Schur superpolynomials.
They generalize the bilinear identity for the standard rectangular-type Schur polynomials found in~\cite{Kir}:
\begin{gather}
\label{BilSchur}
s_{(k^n)}s_{(k^n)}=s_{((k+1)^n)}s_{((k-1)^n)}+s_{(k^{n+1})}s_{(k^{n-1})},
\end{gather}
where as usual $k^n$ means that the part~$k$ is repeated~$n$ times.
This identity can be represented diagrammatically as (with $k=3$ and $n=2$)
\[
\tableau[scY]{&& \\&&\\} \;\times\; \tableau[scY]{&&\\&&\\} =\tableau[scY]{&&&\\&&&\\}
\times
\tableau[scY]{&\\&} \;+\:\tableau[scY]{&& \\&&\\&&}
\times
\tableau[scY]{&&}\,.
\]
This (remarkable) identity is equivalent to the Dodgson's condensation formula (also known as the Desnanot--Jacobi matrix
theorem), which reads (see, e.g.,~\cite{Bres})
\begin{gather}
\label{Dodgson}
\det\big(M_{1}^1\big) \det\big(M_\ell^\ell\big)=\det\big(M^\ell_1\big)\det\big(M^1_\ell\big)+\det(M) \det\big(M_{1,\ell}^{1,\ell}\big),
\end{gather}
where~$M$ is a~$\ell \times \ell$ matrix and $M_i^j$ refers to~$M$ with the~$i$-th line and the~$j$-th column removed.
The relation between~\eqref{Dodgson} and~\eqref{BilSchur} follows from the def\/inition of Schur function in terms of a~determinantal formula.

We now consider generalizations of~\eqref{BilSchur} for the Schur superpolynomials.
Note that tentative proofs along the above lines are bound to fail due to the absence of determinantal formula for
superpolynomials.
Here, a~rectangular diagram can be extended in dif\/ferent ways by the adjunction of circles.
For example for the above diagram $(3^2)$ is generalized by the following four super-diagrams
(for which either $\Lambda^*$ or $\Lambda^\circledast$ is rectangular):
\[
\tableau[scY]{&&\\&&\\\bl\tcercle{}}
\qquad
\tableau[scY]{&&&\bl\tcercle{}\\&&}
\qquad
\tableau[scY]{&&\\&&\bl\tcercle{}}
\qquad
\tableau[scY]{&&&\bl\tcercle{}\\&&\\\bl\tcercle{}}.
\]

\begin{conjecture}
Let $k$, $n$ be integers with $k>1$ and $n>1$.
Let $r', r \in \{k,k-1,0\}$ with $r'>r$ and $\epsilon=\delta_{r,0}$.
We have
\begin{gather}
s_ {(r;k^{n-1+\epsilon})} s_{(k^n)}=s_{(r+1-\epsilon; (k+1)^{n-1+\epsilon})} s_{((k-1)^n)}+s_{(r;k^{n+\epsilon})}s_{(k^{n-1})},
\label{SBil1}
\\
s_ {(r';k^{n-1})} s_{(r; k^{n-1+\epsilon})}=s_{(r'+1,r+1-\epsilon; (k+1)^{n-2+\epsilon})} s_{((k-1)^n)}+s_{(r',r;k^{n-1+\epsilon})}s_{(k^{n-1})},
\label{SBil2}
\\
s_{(k,0; k^{n-1})} s_{(k^n)}=s_{(k+1,0;(k+1)^{n-1})} s_{((k-1)^n)}+s_{(k,0; k^{n})} s_{(k^{n-1})}  .
\label{SBil3}
\end{gather}
\end{conjecture}

Note that in the f\/irst two cases, each identity is a~compact formulation for three identities as there are three ways of
selecting the pair $(r,r')$.
For example, the identity~\eqref{SBil2} with $k=3$, $n=2$ and $r'=3$, $r=2$ reads
\[
\tableau[scY]{&&&\bl\tcercle{}\\&&\\} \;\times\; \tableau[scY]{&&\\&&\bl\tcercle{}\\} =
\tableau[scY]{&&&&\bl\tcercle{}\\&&&\bl\tcercle{}\\}  \;\times\;  \tableau[scY]{&\\&} \;+\:\tableau[scY]{&&&\bl\tcercle{}\\&&\\&&\bl\tcercle{}}
\times
\tableau[scY]{&&}\, .
\]
For the same values of~$k$ and~$n$, the diagrammatic form of~\eqref{SBil3} is
\[
\tableau[scY]{&&&\bl\tcercle{}\\&&\\\bl\tcercle{}} \;\times\; \tableau[scY]{&&\\&&\\}
=\tableau[scY]{&&&&\bl\tcercle{}\\&&&\\\bl\tcercle{}}  \;\times\;  \tableau[scY]{&\\&} \;+\:\tableau[scY]{&&&\bl\tcercle{}\\&&\\&&\\\bl\tcercle{}}
\times
\tableau[scY]{&&}\, .
\]

In addition, similar bilinear identities have been found for almost rectangular-type super-diagrams, namely
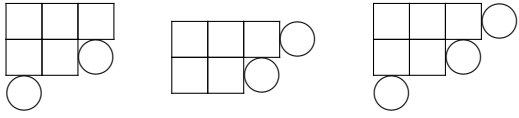
\[
\tableau[scY]{&&\\&&\bl\tcercle{}\\\bl\tcercle{}}
\qquad
\tableau[scY]{&&&\bl\tcercle{}\\&&\bl\tcercle{}}
\qquad
\tableau[scY]{&&&\bl\tcercle{}\\&&\bl\tcercle{}\\\bl\tcercle{}}
\]
for which neither $\Lambda^*$ nor $\Lambda^\circledast$ rectangular.

\begin{conjecture}
We have
\begin{gather}
 s_{(k,k-1; k^{n-2})} s_{(k^n)}=- s_{(k+1,k;(k+1)^{n-2})} s_{((k-1)^n)}+s_{(k-1; k^{n})} s_{(k;k^{n-2})},
\label{SBila}
\\
 s_{(k-1,0; k^{n-1})} s_{(k^n)}=s_{(k-1,0;(k+1)^{n-1})} s_{((k-1)^n)}+s_{(k-1; k^{n})} s_{(0;k^{n-1})},
\label{SBilb}
\\
 s_{(k,k-1;k^{n-2})} s_{(0;k^n)}=s_{(k+1,k,0; (k+1)^{n-2})} s_{((k-1)^n)}+s_{(k-1,0; k^{n})}s_{(k;k^{n-2})},
\label{SBilc}
\\
 s_{(k,k-1,0;k^{n-2})} s_{(k^n)}=s_{(k+1,k; (k+1)^{n-2})} s_{(0;(k-1)^n)}+s_{(k-1,0; k^{n})}s_{(k;k^{n-2})}.
\label{SBild}
\end{gather}
\end{conjecture}

Still for the values $k=3$ and $n=2$, we illustrate the identities~\eqref{SBilc}:
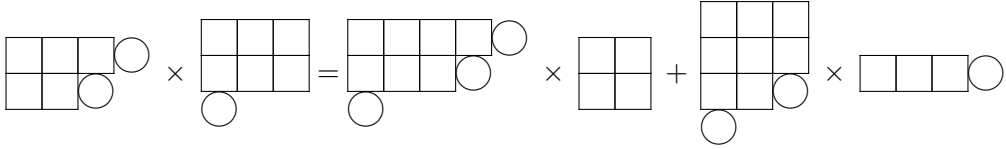
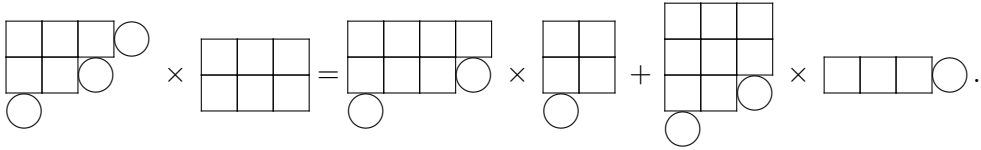
\[
\tableau[scY]{&&&\bl\tcercle{}\\&&\bl\tcercle{}\\} \;\times\; \tableau[scY]{&&\\&&\\\bl\tcercle{}} =
\tableau[scY]{&&&&\bl\tcercle{}\\&&&\bl\tcercle{}\\\bl\tcercle{}}  \;\times\;  \tableau[scY]{&\\&} \;+\:
\tableau[scY]{&&\\&&\\&&\bl\tcercle{}\\\bl\tcercle{}}  \;\times\;  \tableau[scY]{&&&\bl\tcercle{}}
\]
and~\eqref{SBild}:
\[
\tableau[scY]{&&&\bl\tcercle{}\\&&\bl\tcercle{}\\\bl\tcercle{}} \;\times\; \tableau[scY]{&&\\&&\\} =
\tableau[scY]{&&&\\&&&\bl\tcercle{}\\\bl\tcercle{}}  \;\times\;  \tableau[scY]{&\\&\\\bl\tcercle{}} \;+\:
\tableau[scY]{&&\\&&\\&&\bl\tcercle{}\\\bl\tcercle{}}  \;\times\;  \tableau[scY]{&&&\bl\tcercle{}}\, .
\]

\subsection*{Acknowledgements}

We thank Luc Lapointe for useful discussions and critical comments on the manuscript.
We also thank Patrick Desrosiers for his collaboration at the early stages of this project.
This work is supported by NSERC and FRQNT.

\pdfbookmark[1]{References}{ref}
\LastPageEnding

\end{document}